\documentclass[letterpaper,twocolumn,10pt]{article}
\usepackage{hyperref}
\usepackage{usenix-2020-09}

\usepackage[normalem]{ulem}

\usepackage{breakurl}
\usepackage[normalem]{ulem}
\usepackage{endnotes}
\usepackage{url}

\usepackage{tabularx}
\usepackage{pifont}

\usepackage{epsfig,multirow,makecell,color}
\usepackage{amsfonts}
\usepackage{amssymb}
\usepackage{fdsymbol}
\usepackage[linesnumbered,boxed,lined,ruled,vlined]{algorithm2e}
\usepackage{listings}
\usepackage{xcolor}
\usepackage{pifont}
\usepackage{colortbl}
\usepackage{balance}
\usepackage{graphicx}
\usepackage{float}
\usepackage[compact]{titlesec}
\usepackage{wrapfig}
\usepackage[export]{adjustbox}
\usepackage{datetime}
\usepackage{cleveref}
\usepackage{ifsym}
\usepackage{enumitem}
\usepackage{thmbox}
\usepackage[labelfont=bf,font=normal,skip=1pt]{caption}

\usepackage{tikz}

\newcommand{\TODO}[1]{\textcolor{red}{TODO: #1}}

\usepackage{listings}
\usepackage{lstautogobble}

\usepackage{booktabs}
\usepackage{multirow}

\newcommand{\sys}{DuVisor}

\newcommand{\ignore}[1]{}

\begin{document}

\title{{\sys}: a User-level Hypervisor Through Delegated Virtualization}

\author{Jiahao Chen, Dingji Li, Zeyu Mi\thanks{Zeyu Mi (yzmizeyu@sjtu.edu.cn) is the corresponding author}, Yuxuan Liu,
Binyu Zang, Haibing Guan, Haibo Chen\\
{\normalsize \it {Institute of Parallel and Distributed Systems, Shanghai Jiao Tong University}} \\
} 

\date{}
\maketitle

\thispagestyle{empty}

\begin{abstract}
Today's mainstream virtualization systems comprise of two cooperative components: 
a kernel-resident driver that accesses virtualization hardware and a user-level helper process that 
provides VM management and I/O virtualization. 
However, this virtualization architecture has intrinsic issues in both security (a large attack surface) and performance.
While there is a long thread of work trying to minimize the kernel-resident driver by offloading functions to user mode,
they face a fundamental tradeoff between security and performance: more offloading may reduce 
the kernel attack surface, yet increase the runtime ring crossings between the helper process and the driver, and thus more performance cost.

This paper explores a new design called delegated virtualization, which completely separates the control plane (the kernel driver)
from the data plane (the helper process) and thus eliminates the kernel driver from runtime intervention. 
The resulting user-level hypervisor, called {\sys}, can handle all VM operations without 
trapping into the kernel once the kernel driver has done the initialization. 
{\sys} retrofits existing hardware virtualization support with a new delegated virtualization extension to directly handle VM exits, 
configure virtualization registers, manage the stage-2 page table and virtual devices in user mode.
We have implemented the hardware extension on an open-source RISC-V CPU and built a Rust-based hypervisor atop the hardware.
Evaluation on FireSim shows that {\sys} outperforms KVM by up to 47.96\% in a variety of real-world 
applications and significantly reduces the attack surface.
\end{abstract}


\section{Introduction}
\label{sec:intro}

\begin{figure*}
    \centering
    \includegraphics[scale=0.73]{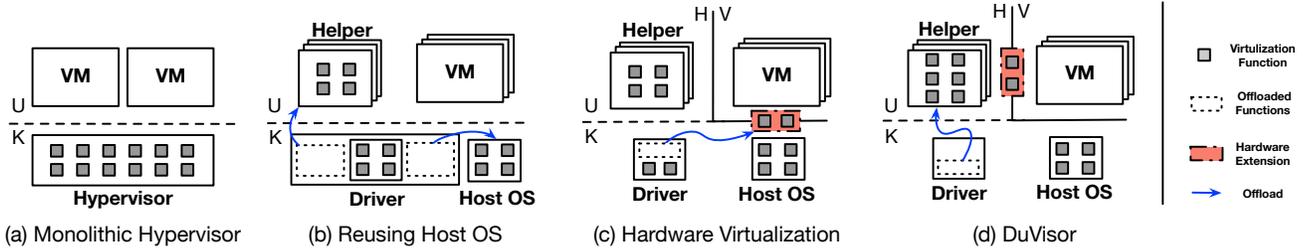}
    \caption{\small{The architectural evolvement of mainstream hypervisors: (a) Stage-1: The monolithic hypervisor puts all virtualization functions in kernel mode; 
    (b) Stage-2: Offloading some functions to a helper process can reuse host OS/management VM that manage hardware resources; 
    (c) Stage-3: Virtualization extensions boost some functions through hardware;
    (d) The {\sys} approach delegates hardware interfaces to user space and eliminates runtime intervention of the host kernel.}}
    \label{fig:hv-comp}
\end{figure*}

System virtualization (or called virtualization for brevity) is a key technique to efficiently run concurrent virtual machines (VM). 
This technique has gone through three rough stages of evolution since its inception (Fig.~\ref{fig:hv-comp}(a)-(c)).
It multiplexed scarce resources of large and expensive mainframe machines in the first stage. 
In the early 1970s, IBM's VM/370 hypervisor, for example, ran on System/370 hardware~\cite{creasy1981origin, gum1983system}. 
All virtualization functions were in kernel mode due to the small code size and high reliability of hypervisors at the time~\cite{seawright1979vm, buzen1973evolution}. 
In the second stage, virtualization was leveraged to achieve high scalability on multiprocessors without major modifications to existing operating systems (OS)~\cite{bugnion1997disco, whitaker2002scale}. 
Mainstream hypervisors started to offload some virtualization functions to user mode, which issue system calls to take advantage of a host OS~\cite{bugnion2012bringing} or a management VM~\cite{barham2003xen}.
Other functions remained in kernel mode, such as instruction emulation~\cite{popek1974formal} and memory virtualization~\cite{waldspurger2003memory}.

The third stage began with the release of hardware extensions (e.g., Intel VMX~\cite{intelvmx}, AMD SVM~\cite{amdsvm}, and ARM VE/VHE~\cite{armmanual}), 
which improved VM performance by running most VM code directly, enabling stage-2 address translation, etc. 
Such extensions shifted more virtualization functions to hardware, further reducing the kernel involvement in virtualization. 
But commercial hypervisors still depends on a kernel-resident driver (e.g., KVM~\cite{kivity2007kvm, dall2014kvmarm}) to access the interface of hardware extensions at runtime.

The third-stage hypervisor architecture popularized for nearly two decades because it significantly reduces hypervisor complexity with acceptable 
virtualization overhead.
However, it still has both security and performance issues.
First, despite the fact that many functions have been moved out of the kernel mode,
the kernel driver remains a large Trusted Computing Base (TCB), leading to a number of security vulnerabilities. 
For instance, KVM, whose code base consists of 57K lines of code (LoC),
has accumulated a total of 127 CVEs over the course of its evolution~\cite{kvmcve}.
Since the driver locates in the kernel mode,
a vulnerability can be maliciously exploited to take control of the host kernel and even gain illegal access to
other VMs' data~\cite{epycescape, cve202122543} or cause denial of service (DoS) attacks~\cite{cve202143056, cve201919332}.

Second, existing hypervisor architecture may incur non-trivial virtualization overhead. 
The kernel driver invokes rich functionalities (e.g., physical memory allocation) implemented by 
the host kernel, but these general functionalities are not well optimized for virtualization. Worse,
some VM exits in the traditional hypervisor are forwarded to
the user-mode management software for processing, causing excessive ring crossings and thus performance overhead~\cite{humphries2021case}.
To mitigate the ring crossing issue, some I/O virtualization functions are put back to kernel mode (like vhost-net~\cite{vhost-net}),
but this design sacrifices security for performance and even leads to VM escape (e.g., V-gHost~\cite{vhostbug}).

A long line of research has tried to minimize the kernel part and deprivilege unnecessary functions to user mode~\cite{steinberg2010nova,
colp2011breaking, murray2008improving, wu2013taming}. Yet, they face a fundamental \emph{tradeoff}:
a smaller kernel driver improves system security, yet offloading more functions may bring more ring crossings as they have to
depend on the kernel to drive the hardware extension~\cite{wu2013taming}.

The root cause of this performance-security tradeoff is the unnecessarily tight coupling of hardware virtualization to kernel mode.
Fortunately, we observe that recent hardware advances can expose physical resources to user mode that used to be managed only by the kernel.
One prominent example of them is that Intel and RISC-V both released user-level interrupts, which allow a user-level process to 
handle physical interrupts~\cite{inteluipi,riscv-privileged}.
Another example is physical memory checking (e.g., RISC-V Physical Memory Protection or PMP~\cite{lee2020keystone}) that limits the 
physical memory range a program can access.
With such recent hardware features, we believe it is time to rethink the design of hypervisor architecture.
To this end, we propose \emph{delegated virtualization}, which allows a user-level hypervisor 
(called DuVisor\footnote{Short for \underline{D}elegated \underline{u}ser-level Hyper\underline{Visor}}) to securely and efficiently control all virtualization functions in user mode (see Fig.~\ref{fig:hv-comp}(d)), 
with a tiny kernel driver only for the initialization of {\sys} and handing of fatal faults. 

We design the delegated virtualization by a novel hardware extension called Delegated Virtualization Extension (DV-Ext). 
DV-Ext mostly reuses existing virtualization extensions with only minor modifications and securely exposes its hardware interface to user mode. 
With the new extension, the user-level hypervisor is able to handle runtime VM operations without trapping into the host kernel. 
Specifically, {\sys} directly utilizes DV-Ext's registers and instructions in user mode to serve runtime VM exits caused by sensitive instructions, 
stage-2 page faults, and I/O operations. The host kernel is the control plane~\cite{belay2014ix, peter2015arrakis}, 
waking up occasionally only to extend physical resources for {\sys} and handle fatal faults such as {\sys} illegally accessing another VM's memory 
regions (a {\sys} process supports one VM).

{\sys} efficiently provides different virtualization functions in user mode with strong security guarantees.
For CPU virtualization ($\S$~\ref{subsec:vmexit}), a {\sys} process creates a dedicated thread (vthread) for each virtual CPU (vCPU), and the vthread makes use of DV-Ext to handle this vCPU's VM exits in user space. 
All {\sys} threads are managed by the host kernel scheduler, which allows the host kernel to decide how to consume physical CPU resources.
For memory virtualization ($\S$~\ref{subsec:memory_S2PT}), {\sys} configures a stage-2 page table for its VM and processes
stage-2 page faults in user mode via allocating physical memory in user mode.
The physical memory range used by {\sys} is restricted by the host kernel and DV-Ext via hardware physical memory checking. 
For I/O virtualization ($\S$~\ref{subsec:intr_virt}), para-virtualized (PV) backend drivers in user mode directly communicate with their frontends in VMs. 
{\sys} further boosts I/O performance by using user-level inter-processor interrupts (UIPI) to completely bypass the host kernel when sending notifications to its VM.

We have implemented DV-Ext based on RISC-V Rocket CPU using FPGAs. 
DV-Ext can be easily implemented by reusing existing hardware features, including virtualization extension (H-Ext) and user-level interrupts (N-Ext). 
Therefore, DV-Ext only costs 420 lines of Chisel code.
Based on DV-Ext, we use Rust to build {\sys}, and the code size is about 8K LoC. 
We also extend the Linux kernel v5.10.26 with a kernel driver to cooperate with {\sys} by adding 362 LoC.
Performance evaluation on architectural operations and real-world applications show that {\sys} outperforms KVM by up to 47.96\%.

The contributions of the paper are:
\begin{itemize}
    \item We propose delegated virtualization to break the security/performance tradeoff faced by traditional hypervisors.
    \item We design a user-level hypervisor that serves VMs without involving the kernel driver at runtime.
    \item We implement the hardware extension on RISC-V and build a Rust-based hypervisor, both of which will be publicly available~\footnote{https://github.com/IPADS-DuVisor}.
    \item We evaluate {\sys} on AWS F1 FPGAs using cycle-accurate Firesim~\cite{karandikar18firesim} with a suite of real-world applications.
\end{itemize}

\section{Background and Motivation}
\label{sec:bg}

\subsection{Hardware Virtualization Extension}
\label{subsec:h_ext}

Despite minor differences in hardware interfaces, today's virtualization extensions~\cite{intelvmx, amdsvm, armmanual, riscv-hext} provide comparable functionalities, 
such as selective trapping of sensitive instructions, stage-2 address translation, interrupt virtualization etc.
They all demand that their interface be accessed in host kernel mode. 
We take RISC-V H-Ext~\cite{riscv-hext} (see Fig.~\ref{fig:hv-comp}(c)) as an example to explain existing hardware extension. 
H-Ext has two special modes, which are orthogonal to existing privilege levels (U and K for user and kernel respectively~\footnote{Even though the kernel mode in RISC-V is called ``supervisor'' (S) mode, we still call it kernel mode.}). 
The H mode is for the hypervisor while VMs run in V mode.
Only the kernel level in H mode is capable of using virtualization interface like starting/resuming a VM, installing a stage-2 page table (S2PT), and injecting
virtual interrupts. 
So a helper process (like QEMU~\cite{qemu}) must invoke system calls (\textit{ioctl}) to request a kernel driver (KVM) to control VMs. 
The hardware wakes up the driver to handle it when VM exits trigger,
most of which can be served directly by the driver without switching to the helper. 
Take a stage-2 page fault (S2PF) as an example, the driver invokes the host kernel to allocate a physical page and inserts a new address mapping to 
the VM's S2PT, which records mappings from guest physical address (GPA) to host physical address (HPA).
Some VM exits, such as MMIO trapping, cannot be handled by the driver and should be forwarded to the user-level helper for emulation.

\subsection{Hypervisor Architecture Issues}
\label{subsec:flaws}

This section explains the tradeoff faced by today's hypervisor architecture via QEMU/KVM~\footnote{Type-1 hypervisors like Xen~\cite{barham2003xen} and VMware ESXi~\cite{vmwareesxi} use the split architecture as well: user-level processes to run VMs and the kernel-level software to drive hardware extension and resources.}. 
On the one hand, the traditional hypervisor relies on the KVM driver, which has a large TCB and invokes rich kernel APIs that are not well optimized for virtualization.
On the other hand, while shifting most of the driver to user space decreases its TCB, 
the deprivileged execution will have a higher cost owing to the more frequent kernel's involvement in each VM exit handling~\cite{humphries2021case, steinberg2010nova, wu2013taming}.

\subsubsection{Conventional Hypervisors}

\vspace{3mm} \noindent \textbf{Huge TCB and Weak Isolation.} 
A conventional hypervisor (or the kernel driver) has weak security and fault isolation, making it the system's single point of failure.
An adversary could exploit vulnerabilities to target not only the hypervisor, 
but the entire system, including the host kernel and all other cloud tenants' VMs.
Table~\ref{tab:virtual-cve} shows that even for KVM that is just a kernel driver, it has a large TCB (57K LoC on x86-64) and a number of vulnerabilities.
There have been 127 disclosed vulnerabilities of KVM~\cite{kvmcve} since 2008, 
62.99\% of which leak information from the host OS, result in DoS on the host~\cite{shi2017deconstructing, cve202143056, cve201919332} or even allow a guest to escalate privilege~\cite{epycescape, vhostbug, cve202122543, scavenger, 3dredpill, backtofuture, cve20197221}.
Other hypervisors have the same isolation issue. There have also been 423 and 77 CVEs for Xen~\cite{xencve} and VMware ESXi~\cite{vmwarecve} since 2007 and 2008 respectively.
76.36\% and 37.66\% of Xen's CVEs and VMware ESXi's CVEs enable the guests to attack the host kernel.

\begin{table}[]
    \centering
    \resizebox{\columnwidth}{!}{%
    \begin{tabular}{@{}l|crrrr|r|c@{}}
    \toprule
    \multirow{2}{*}{\textbf{Hypervisor}} & \multirow{2}{*}{\textbf{Total}} & \multicolumn{3}{c}{\textbf{Host}}                                                                    & \multicolumn{1}{c|}{\multirow{2}{*}{\textbf{Other}}} & \multirow{2}{*}{\textbf{LoC}} & \multicolumn{1}{l}{\multirow{2}{*}{\textbf{Starting Years}}} \\ \cmidrule(lr){3-5}
                                         &                                 & \multicolumn{1}{c}{\textbf{PE}} & \multicolumn{1}{c}{\textbf{DoS}} & \multicolumn{1}{c}{\textbf{DL}} & \multicolumn{1}{c|}{}                                &                               & \multicolumn{1}{l}{}                                         \\ \midrule
    KVM                                  & 127                             & 17                              & 53                               & 10                              & 47                                                   & 57K                           & 2008                                                         \\
    Xen                                  & 423                             & 109                             & 189                              & 25                              & 100                                                  & 302K                          & 2007                                                         \\
    VMware                               & 77                              & 13                              & 6                                & 10                              & 48                                                   & -                             & 2008                                                         \\ \bottomrule
    \end{tabular}
    }
    \caption{\small{CVE analysis of KVM~\cite{kvmcve}, Xen~\cite{xencve} and VMware ESXi~\cite{vmwarecve}. 
    \textit{Host} stands for the vulnerabilities that could lead to an attack on the host kernel.
    \textit{PE}, \textit{DoS}, and \textit{DL} stand for privilege escalation, denial of service and data leakage.
    \textit{LoC} shows their lines of code on x86-64.
    (VMware's code size is unclear due to the fact that it is proprietary.)
    \textit{Starting Year} stands for the earliest year in which the vulnerabilities were exposed.}}
    \label{tab:virtual-cve}
\end{table}

\vspace{3mm} \noindent \textbf{Suboptimal Internal Mechanism in Kernel.} 
KVM is designed to call existing host kernel functions to avoid reinventing the wheel.
However, their implementation can be used in general scenarios but incurs non-negligible performance impacts for virtualization.
One typical example is the S2PF handling procedure. 
When an S2PF occurs, the CPU control flow traps from a VM into KVM, which invokes Linux to allocate physical memory before adding mappings to this VM's S2PT.
The memory allocation function comprises complex logic for versatility.
For example, this function needs to use the \textit{memslot} abstraction to translate a GPA to the corresponding HPA,
but the memslot involves the use of RCU mechanism, lock dependency checking, reference count checking, etc.
So this physical memory allocation for the fault address accounts for 2,939 cycles (57.32\%) of the S2PF handling procedure in KVM (more details in $\S$~\ref{subsec:microbenchmarks}).

\subsubsection{Deprivileged Execution}

\begin{table}[]
    \centering
    \scriptsize{
        \begin{tabular}{@{}l|lll@{}}
            \toprule
            \textbf{Platform} & \textbf{Kernel} & \textbf{User} & \textbf{Total} \\ \midrule
            ARM               & 4,323           & 1,596         & 5,919          \\
            x86-64            & 2,415           & 1,704         & 4,119          \\ \bottomrule
        \end{tabular}
    }
    \caption{\small{Breakdown of an MMIO read in QEMU/KVM on both ARM and x86-64.
    \textit{Kernel} represents the cycles spent on the in-kernel transfer operations. 
    \textit{User} is the cycles of the I/O emulation and the consumption of VM entry/exit.}}
    \label{tab:long_path_breakdown}
\end{table}


Deprivileging most parts of the kernel driver to user space leads to frequent communication with a long path between the VM and the user-level hypervisor.
Since the hypervisor has to utilize the host kernel to drive the hardware virtualization extension,
more interactions between VM and the deprivileged hypervisor involve the host kernel, resulting in larger performance overhead.
To understand the cost brought by the kernel involvement,
we break down the handling procedure of an MMIO read operation in QEMU/KVM show the VM-VMM communication cost and find out that 73.04\% and 58.63\% of CPU cycles account for in-kernel transfer operations on ARM and x86 respectively (Table~\ref{tab:long_path_breakdown}).
Minimizing the host kernel part by deprivileging more kernel functionalities to user space~\cite{wu2013taming, steinberg2010nova} will experience much more overhead, because each VM exit has to be forwarded by the kernel to user space, and vice versa.

\section{{\sys} Overview}
\label{sec:overview}

In this paper, we present \textbf{delegated virtualization} to break the tradeoff facing the conventional hypervisor architecture. 
It allows all virtualization functions to be delegated to user mode so that we can build a user-level {\sys}.
We have two design goals for {\sys}. First, the runtime interactions of a guest and its hypervisor should not involve the host kernel.
Second, the vulnerabilities of {\sys} should be confined in user space, which does not affect other guest VMs or the host kernel.
We obtain these properties via redesigning the hardware interface of the virtualization extension and taking the control plane/data plane separation idea~\cite{peter2015arrakis, belay2014ix}. 
The host kernel works as the control plane and is removed
from all runtime interactions (data path) between {\sys} and its VM.

\begin{figure}
    \centering
    \includegraphics[width=0.43\textwidth]{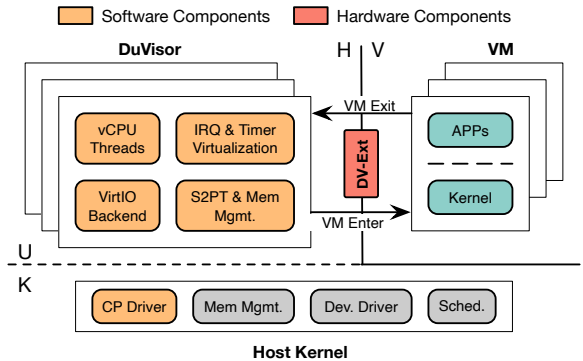}
    \caption{\small{The architecture overview of {\sys}. DV-Ext is Delegated Virtualization Extension and \textit{CP} stands for control-plane.}}
    \label{fig:arch}
\end{figure}

The architecture, which consists of three major components, is shown in Fig.~\ref{fig:arch}.
\textbf{First}, Delegated Virtualization Extension (DV-Ext) should be installed on the hardware ($\S$~\ref{sec:hardware}).
It empowers the host kernel to decide whether or not to delegate hardware virtualization functions to hypervisor user mode, so that {\sys} can directly control a VM's behaviors and handle VM exits 
without trapping into the host kernel.

\textbf{Second}, a {\sys} process utilizes hardware instructions to manage and serve an unmodified VM without trapping into the host kernel ($\S$~\ref{sec:software}).
Most VM exits of its corresponding VM can be handled by {\sys} directly, including S2PF. {\sys} also configures an S2PT for its VM.
So it adds a new mapping entry into the S2PT if an S2PF happens due to missing a page mapping.
A single VM is served by a {\sys} process.
If multiple VMs are needed, separate {\sys} processes should be created for each VM.
{\sys} creates multiple threads for different vCPUs (we name these threads as \textit{vthreads}), like traditional hypervisors. 
It also provides virtual physical memory, PV I/O devices, virtual interrupt and timer for this VM. 
{\sys} can not only depend on the host kernel to manage external devices like storage media and network cards,
but also control devices in user mode by DPDK~\cite{dpdk} to boost I/O virtualization.

\textbf{Lastly},
a tiny control-plane (CP) driver enhances the host kernel to become the control plane for {\sys} ($\S$~\ref{sec:impl}). 
The CP driver has the privilege to determine whether or not to delegate hardware virtualization functions to a process. 
It allocates hardware resources for {\sys} during its startup phase, 
creates the {\sys} process, and handles emergency cases such as a page fault triggered by an illegal stage-2 address mapping.
We rely on the host kernel to schedule {\sys} and its VM. 
Recent advances in user-defined scheduling policies~\cite{humphries2021ghost} can be applied to {\sys} in the future.

\vspace{3mm} \noindent \textbf{Assumptions and Threat Model.} 
We assume that the hardware and DV-Ext are correctly implemented and trusted. 
The host kernel is trusted as well, even though it may contain vulnerabilities. 
We consider that the attacker can take full control of a guest VM and further compromise 
{\sys} by exploiting its vulnerabilities. Therefore, {\sys} is untrusted by the host kernel.

\section{Delegated Virtualization Extension}
\label{sec:hardware}

\begin{table}[]
    \centering
    \scriptsize{
    \begin{tabular}{p{2mm}|p{3mm}p{10mm}l}
    \toprule
    \multicolumn{1}{c|}{\textbf{Type}}                          & \textbf{Mode}                     & \multicolumn{1}{l}{\textbf{Name}} & \textbf{Description}                                                      \\ \hline
    \multirow{7}{*}{\textbf{Registers}}                         & \multirow{5}{*}{HU}      & hu\_er                   & VM exit reason                                                   \\
                                                       &                          & hu\_einfo                    & Additional information about a VM exit \\
                                                       &                          & hu\_vitr                     & Virtual interrupt number to be inserted                \\
                                                       &                          & hu\_vpc                     & IP address of a faulted vCPU                                     \\
                                                       &                          & hu\_ehb                    & Base address of the VM exit handler                                      \\
                                                       &                          & hu\_vcpuid                  & The vCPU ID running on a physical core                   \\ \cline{2-4}
                                                       &                          & h\_enable                   & Turn on DV-Ext                                  \\
                                                       &   \multirow{1}{*}{HS}    & h\_deleg                   & Delegate VM exits to HU-mode                         \\ 
                                                       &                          & h\_vmid                  & The VM ID running on a physical core                   \\\hline
    \multicolumn{1}{c|}{\multirow{2}{*}{\textbf{Instructions}}} & \multirow{2}{*}{HU}      & HURET                     & Resume the vCPU execution         \\
    \multicolumn{1}{c|}{}                              &                          & HUSUIPI                  & Send a UIPI to a physical core                                   \\ \bottomrule
    \end{tabular}
    }
    \caption{\small{The registers and instructions added (or modified) by DV-Ext. The lowercase and uppercase names stand for registers and instructions. The registers starting with ``hu'' are accessible in HU-mode while those starting with ``h'' can only be accessed in HS-mode.}}
    \label{tab:huext-interface}
\end{table}

Recent hardware advances reveal a trend that disentangles the management of physical resources from their protection. 
A user-level process can utilize physical resources freely within a predefined range configured by the OS kernel, 
but unauthorized use outside the range will trigger an exception to wake up the kernel.
These hardware features shed light on a new scenario where the kernel can expose hardware resources to user mode for efficiency while retaining protection. 

Inspired by such hardware trend, we present a novel interface of hardware virtualization: Delegated Virtualization Extension (\textbf{DV-Ext}), whose registers and instructions are shown in Table~\ref{tab:huext-interface}.
DV-Ext empowers the hypervisor kernel mode (HS-mode) to decide whether or not some hardware virtualization functions (their corresponding registers and instructions) can be delegated to the hypervisor user mode (HU-mode).
Afterwards, HU-mode software can directly control the delegated virtualization functions and resources without relying on the host kernel.
Through reusing recent hardware features with minor modifications, DV-Ext prohibits a hostile HU-mode hypervisor from maliciously manipulating delegated hardware resources.


\subsection{HU-mode Registers and Instructions}
\label{subsec:humode}

One straightforward method to enable HU-mode is to allow the HU-mode to access all virtualization registers and instructions. 
But it may disclose unnecessary information to HU-mode, making DV-Ext difficult to implement. 
We observe that most registers are only configured during VM initialization and rarely touched at runtime. 
So these registers can be regarded as control-plane registers and will not be exposed to HU-mode.
The data-plane registers, on the other hand, are the remaining registers that are frequently used at runtime. 
DV-Ext permits these data-plane registers to be accessed in HU-mode, which are shown in Table~\ref{tab:huext-interface} as registers beginning with ``hu''. 
These data-plane registers can be accessed in HU-mode only when HS-mode turns on DV-Ext by configuring \textit{h\_enable}.
The data-plane registers are classified into two categories.
The first category records the VM information for VM exits, such as \textit{hu\_er} and \textit{hu\_einfo}.
The hypervisor reads these registers for handling VM exits.
The second category affects the runtime behaviors of the hypervisor or VMs.
For instance, a hypervisor configures \textit{hu\_vitr} to inject a virtual interrupt to a vCPU.

Today's hardware can delegate physical interrupts and exceptions to user mode (e.g., RISC-V user-level interrupt extension or N-Ext).
DV-Ext retrofits such feature to support delegatable VM exits (DVE) that can be delegated to HU-mode.
DVE is configured by the host kernel by setting up the \textit{h\_deleg} register, whose individual bit controls the delegation of one specific type of VM exit.
The host kernel in HS-mode can delegate S2PF and sensitive instruction faults (such as WFI) to HU-mode by setting the corresponding bits in \textit{h\_deleg}.
When a DVE happens, the hardware locates a hypervisor handler whose address is specified in \textit{hu\_ehb}.
DV-Ext also provides the \textit{HURET} instruction for HU-mode to resume the VM execution, whose entry point is stored in the \textit{hu\_vpc} register. 

\subsection{Dynamic HPA Checking}
\label{subsec:vbit}

DV-Ext does not expose the register holding the base address of an S2PT to HU-mode since it is rarely modified after a VM is booted.
But it still allows a user-level hypervisor to freely configure S2PT address mappings in HU-mode. One natural challenge is how to prevent the hypervisor from maliciously configuring
the S2PT to access arbitrary HPA.
Today's hardware usually contains a mechanism to dynamically check physical memory range according to hardware register configurations, such as ARM TrustZone and RISC-V Physical Memory Protection (PMP).
DV-Ext leverages such mechanisms to restrict the physical memory range that the hypervisor and its VM can touch.

The physical memory checking (PMC) mechanism provides a set of per-core registers that specify different physical memory regions accessible to software. 
In our current implementation, the largest number of different regions for a physical core is 64.
PMC examines the length and attributes of a physical memory access according to these registers and triggers an exception if detecting illegal physical addresses. 
Since the dynamic checking is implemented by comparing offset, its overhead is negligible.
DV-Ext slightly extends an existing PMC mechanism to make it only work for HPA translated from guest physical addresses (GPA).
Specifically, DV-Ext adds a ``Virtualization'' (V) bit to each range register to indicate that it is valid only for HPA from the stage-2 address translation.

\subsection{Boosting Interrupt Virtualization}
\label{subsec:uipi}

DV-Ext proposes a user-level notification mechanism to boost virtual interrupt injection, which is built based on existing user-level interrupt mechanism~\cite{inteluipi,riscv-privileged}.
An HU-mode program in one physical core can invoke \textit{HUSUIPI} instruction to send a user-level inter-processor interrupt (UIPI) to another physical core.
If the receiving core is running in V-mode, this UIPI causes a VM exit to the HU-mode.
The sending core should write its VMID and VCPUID to two registers (\textit{h\_vmid} and \textit{hu\_vcpuid}) 
and specify the receiver's VCPUID as the operand for \textit{HUSUIPI}. 
The hardware will check the ID information when delivering a UIPI.
An illegal operand will trigger a fault into the HS-mode to wake up the host kernel.
\section{{\sys} Design}
\label{sec:software}

\subsection{Handling VM Exits}
\label{subsec:vmexit}

\begin{figure}
    \centering
    \includegraphics[scale=1.7]{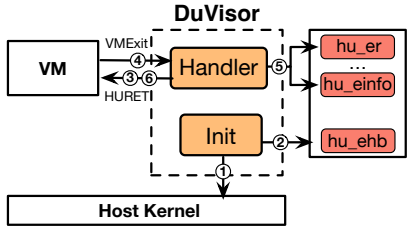}
    \caption{\small{The general workflow of how {\sys} handles VM exits in HU-mode.}}
    \label{fig:handling-vmexit}
\end{figure}

Equipped with DV-Ext, a {\sys} process is able to serve a VM in HU-mode without trapping into the host kernel. 
Before using DV-Ext, {\sys} should invoke a system call (ioctl) to ask the CP driver to turn on DV-Ext (write \textit{h\_enable}) and delegate VM exits to HU-mode.
After that, {\sys} can freely utilize DV-Ext registers and instructions.
Fig.~\ref{fig:handling-vmexit} shows an example of how {\sys} handles VM exits. {\sys} first installs a VM exit handler by writing its address 
into \textit{hu\_ehb}. 
Before running the first vCPU, {\sys} sets up the VM execution environment, including configuring an S2PT and preparing virtual I/O devices.
It also initializes general-purpose registers and system registers for the vCPU.

{\sys} invokes an \textit{HURET} instruction to enter V-mode and starts to run the guest's code.
When a VM exit like a sensitive instruction (\textit{WFI}) or an S2PF triggers, hardware directly calls the handler specified 
in \textit{hu\_ehb}, which then handles this VM exit by reading corresponding HU-mode registers.
For example, it reads \textit{hu\_er} to check whether this is an S2PF and gets fault address by accessing \textit{hu\_einfo}.
After finishing processing the VM exit, {\sys} resumes the VM by invoking \textit{HURET} again.

Even though the host kernel is not involved in the runtime interactions between {\sys} and the VM, 
it is still responsible for scheduling all processes. Therefore, physical timer interrupts are not delegated to {\sys}.
These interrupts cause VM exits into the host kernel, giving it the opportunity to reclaim control of the physical cores assigned to {\sys} periodically. 
When a timer interrupt happens if a vCPU is running, the VM exit traps into the host kernel, which
saves the current state of the vCPU before switching to another process. The kernel will restore the vCPU's states before resuming it again.

\subsection{Restricted Memory Virtualization}
\label{subsec:memory_S2PT}

A hypervisor needs to create an S2PT for its VM, which records mappings from GPAs to HPAs. 
As we have analyzed in $\S$~\ref{sec:bg}, the traditional kernel driver like KVM relies on the complex memory management in Linux to manage S2PTs, leading to runtime overhead.
In contrast, {\sys} builds the S2PT and memory in HU-mode directly, simplifying the complexity of memory virtualization.
Since an S2PT records mappings from GPA to HPA, {\sys} should see HPA in HU-mode.
Hence, it asks the host kernel to pre-allocate a contiguous pinned memory region and return the base physical addresses (HVA and HPA) of this region.
The memory region is pinned by the host kernel so that the region's pages will not be swapped out during runtime. 
Each S2PF wakes up {\sys} to allocate a free physical page from the memory region and 
add this page into the VM's GPA space by modifying the S2PT.

\vspace{3mm} \noindent \textbf{Restricting Physical Memory Regions in HU-mode.}
However, it is dangerous for an untrusted process to modify an S2PT because it could maliciously map another VM's physical memory pages to 
a malicious VM, which then reads (or alters) sensitive data in the victim VM's memory and passes the data to {\sys}. 
Worse, the rogue VM can even read and modify the host kernel memory. 
This threat can be mitigated by a straightforward approach, which requires {\sys} to invoke system calls to 
ask the host kernel to check this table before installation. 
Specifically, {\sys} should manage a fake S2PT instead of the really used one. 
The host kernel maintains a table to track page ownership and checks each page table mappings
passed by {\sys} before synchronizing to the real S2PT.
Although this approach sounds reasonable, it frequently involves the kernel involvement at runtime, leading to significant cost for memory-intensive workloads. 
Moreover, it complicates memory virtualization in the CP driver.

We take a different approach that allows {\sys} to freely modify the real S2PT in HU-mode without trapping into kernel mode. 
We propose to take advantage of the existing hardware PMC mechanism as we have introduced in $\S$~\ref{subsec:vbit}. 
Equipped with PMC, we can optimistically allow {\sys} to freely configure its S2PT.
The MMU automatically checks whether HPAs accessed by the VM exceed a predefined range limit of the allocated physical memory regions. If so, the MMU triggers a fault to wake up the CP driver.
This design completely eliminates the stage-2 memory management module in the CP driver.

However, the existing PMC mechanism is not specially designed to restrict memory accesses from VMs. 
It checks all physical memory used by the current physical core.
Therefore, it may even restrict the host kernel and the {\sys} process from using physical addresses exceeding the register ranges,
which is a severe limitation given that the host kernel and {\sys} can possibly access all physical memory space. 
That is the reason why DV-Ext slightly 
modifies the existing PMC mechanism by adding a V bit in these registers as we have introduced in $\S$~\ref{subsec:vbit}.
This bit indicates whether or not this region takes effect in V-mode. 
The MMU only checks HPA translated from S2PT or S2PT page table pages if the bit is set.

\begin{figure}
    \centering
    \includegraphics[scale=1.3]{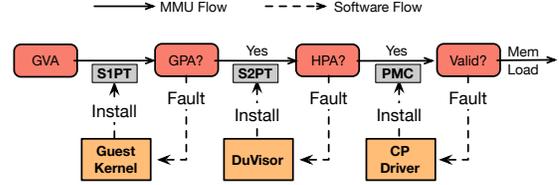}
    \caption{\small{The procedure of PMC checking.}}
    \label{fig:pmp-vbit}
\end{figure}

\vspace{3mm} \noindent \textbf{Put the Pieces Together.}
Before configuring an S2PT, {\sys} should invoke system calls to request a contiguous physical memory region. The CP driver 
then set PMC registers to mark the start and end addresses of this region. It also enables the \textit{V} bit for this region to indicate 
the MMU only checks physical addresses from V-mode (Please note that the S2PT pages are also guarded). 
The CP driver then returns the base physical address of the region and its region size. 
Afterwards, {\sys} builds an S2PT for its VM.
When running out of physical memory, {\sys} can invoke system calls to request more memory regions from the CP driver.

Fig.~\ref{fig:pmp-vbit} shows an example of how a GVA is translated into an HPA and finally used to load memory content.
During runtime, if the MMU cannot find mappings or lack enough permissions for one translation request for S1PT or S2PT,
it will generate a page fault to invoke the corresponding handler in the guest kernel or {\sys}. After the stage-2 translation, the output HPA will 
be checked by the MMU before loading memory contents. If the HPA exceeds the region limit, a page fault is triggered and wakes up the host kernel 
to handle this situation.

\subsection{I/O and Interrupt Virtualization}
\label{subsec:intr_virt}

{\sys} provides PV (e.g., virtio) and emulated (e.g., tty) I/O devices for its VM.
During the initialization of a VM, {\sys} creates dedicated I/O thread(s) for each virtual device.
These threads are typically in charge of dealing with I/O requests for the VM and interacting with host I/O devices.
For example, to receive network packets from the host physical NIC and insert them into the guest VM,
one RX thread is created for the RX queue of a virtio network device.
The design of {\sys} can be combined with a kernel-bypass virtio approach like vhost-user to further boost I/O virtualization.
Specifically, the RX thread keeps polling on the NIC in HU-mode to process incoming network packets and
notifies the guest VM via injecting virtual interrupts.

In a traditional hypervisor, if I/O threads intend to inject virtual interrupts to a vCPU, it invokes a system call 
to send an \textit{eventfd} to a vCPU. Then the host kernel injects a virtual interrupt to the vCPU.
To ensure that the guest VM receives and handles interrupts in a timely manner,
we present an efficient user-level notification mechanism based on a UIPI technique, which requires no kernel involvement and greatly reduces network latency overhead in traditional hypervisors.
In fact, the UIPI technique has already been presented by Intel~\cite{inteluipi}. We slightly extend this technique to fit the virtualization scenario.

The notification mechanism in {\sys} consists of two cases. In the first case where the target vCPU is not running, the sending thread just writes the interrupt information
in the vCPU's state area. Before the vthread resumes the vCPU execution, it will inject the virtual interrupt into the vCPU via loading the state area information into the \textit{hu\_vintr} register.
The second case is that the vCPU is running. The sending thread first writes the interrupt information to the target vCPU state area
and then invokes \textit{HUSUIPI} to send a UIPI to the target vCPU. 
Since a UIPI is a physical interrupt, it will shoot down the running vCPU to cause a VM exit into {\sys}. 
Afterwards, the corresponding vthread injects a virtual interrupt into the vCPU and resumes its execution.

Nevertheless, a malicious process may exploit \textit{HUSUIPI} to keep sending UIPIs, and the frequent VM exits could lead to a DoS attack, which seriously disturbs other VMs.
So the DV-Ext forbids HU-mode from explicitly specifying any target physical core it intends to send UIPIs to. 
Instead, it can only send UIPIs by specifying a VCPUID, which is written in \textit{hu\_vcpuid} by each vthread before running the corresponding vCPU.
Only the receiver's VCPUID matches the HUSUIPI's operand and the sender and receiver share 
the same VMID, the hardware will inject the UIPI to the receiver's core.

\begin{figure}
    \centering
    \includegraphics[scale=1.3]{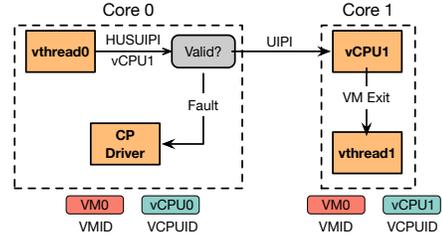}
    \caption{\small{The procedure of sending a UIPI from core 0 to core 1 in HU-mode.}}
    \label{fig:uipi}
\end{figure}

Fig.~\ref{fig:uipi} depicts how {\sys} securely supports UIPI. 
Before {\sys} starts to run, the CP driver initializes a VMID and writes this value to \textit{h\_vmid} registers in each physical core {\sys} will run (core 0 and 1 in this figure). 
The vthreads also specify VCPUIDs into their corresponding \textit{hu\_vcpuid} registers.
During runtime, the sending thread invokes HUSUIPI and provides the target VCPUID as the operand. 
The hardware will generate a physical UIPI to the target CPU and cause a VM exit when the vCPU is running.
Then the VM exit handler is called, whose exit reason is UIPI.
If the thread tries to send a wrong VCPUID or the sending thread does not have the same VMID as the receiver, 
the invocation of HUSUIPI will trigger an exception and wake up the CP driver to handle this issue.

\section{Implementation}
\label{sec:impl}

\subsection{DV-Ext Implementation}
\label{subsec:huext_impl}

We choose the RISC-V platform to implement DV-Ext
since it has rich open-sourced implementations of system-on-chip (SoC)~\footnote{Please note that DV-Ext is not restricted to RISC-V and can be applied to
other architectures as well.}.
We use a 5-stage in-order scalar processor (RISC-V Rocket Core~\cite{rocket-chip}), whose configuration is: 16KB L1 ICache, 16KB L1 DCache, 512KB shared L2 cache, and 16GB external DRAM.

DV-Ext does not require extensive modifications to the CPU hardware to implement these data plane registers and instructions.
These registers are just aliases of existing HS-mode registers.
For example, \textit{hu\_er} and \textit{hu\_einfo} are aliases of \textit{ucause} and \textit{utval} from RISC-V N-Ext (i.e., the user-level interrupt extension) and \textit{HURET} is implemented based on 
N-Ext's \textit{URET}.
Therefore, most architectural implementation for these registers and instructions can be reused.
The RISC-V N-Ext also supports the delegation of exceptions from the kernel to the user space. 
DV-Ext enhances this feature to support DVE so that VM exits could directly trap to HU-mode and get handled by {\sys} directly.

Our DV-Ext implementation adds 420 lines of Chisel to extend the existing H-Ext implementation~\cite{h-ext-pr}.
Moreover, we implement a part of the N-Ext with 70 lines of Chisel,
as its implementation is not yet available but is necessary to
support user-level exception handling (e.g., VM exits from V-mode to HU-mode).
14 lines of Chisel are added to
extend RISC-V PMP for physical memory restriction and enable UIPI for user-level notification mechanism.

\subsection{Software Implementation}
\label{subsec:software_impl}

Our prototype system of {\sys} consists of 8K LoC
(6,732 lines of Rust, 1,632 lines of C and 163 lines of assembly).
In the implementation of CPU and memory virtualization, 4,706 lines of Rust are written
for the main logic, such as VM exit handling and virtual interrupt emulation.
Another 163 lines of C and assembly are used for accessing RISC-V CSRs (Control and Status Register) and
invoking libc routines such as system calls.

To reduce the code effort, our prototype of {\sys} reuses a portion of the I/O backend implementation
(i.e., virtio block and virtio network) from the kvmtool and add 68 lines of C to glue the Rust and C.
We apply our design (e.g., user-level notification mechanism) to its implementation
as well as make some optimizations. Since there is no available DPDK support for RISC-V platforms currently,
we implement a kernel-bypass NIC driver whose code size is 623 in the network backend to achieve similar performance to OVS-DPDK.

We write a tiny Linux kernel module to work as the CP driver, which works
as the control plane for {\sys}. The driver has 317 LoC.
CP driver provides an \textit{ioctl} system call for {\sys} to request services.
First, the CP driver allows {\sys} to turn on DV-Ext for one process before using HU-mode and other features.
Second, the driver allocates contiguous physical memory regions for {\sys} and prevents them from being swapped out.
Then it configures PMP registers to restrict the HPA range that {\sys} can use during runtime. The host kernel should also 
have a PMP fault handler that destroys the fault process.
Lastly, the CP driver initializes a VMID for {\sys} to be used by UIPI.
We also modify the context switch logic (45 LoC) in the host kernel to save and store the DV-Ext registers if the process has enabled DV-Ext.
\section{Security Analysis}
\label{sec:security}

In this section, we analyze the overall system security of {\sys} from the perspective of attackers. 
We also combine CVE cases to illustrate the security advantages of {\sys}.

\vspace{3mm} \noindent \textbf{Attack from Guest to Host Kernel.} 
Since most VM exits directly trap to {\sys} in HU-mode, the host kernel's attack surface is minimized.
A malicious guest may exploit the vulnerabilities, such as CVE-2021-29657~\cite{epycescape} and V-gHost~\cite{vhostbug}, 
to escape from a VM and compromise the traditional hypervisor and the host kernel directly.
In {\sys}, such attacks could only compromise a user-level process.
The attacker cannot further take over the host kernel without additional kernel vulnerabilities that are beyond the consideration of this paper.
Moreover, the host kernel can apply sandboxing techniques (such as seccomp~\cite{seccomp}) to further restrict the {\sys} process.

\vspace{3mm} \noindent \textbf{Attack from Guest to Guest.} 
A malicious guest could attack the other guests on traditional hypervisors~\cite{cve202029480, cve20163159, cve20163158, cve20158555}.
For example, CVE-2020-29480~\cite{cve202029480} allows the guest to leak information of the other guests via the shared hypervisor.
Also, CVE-2016-3159~\cite{cve20163159} and CVE-2016-3158~\cite{cve20163158} allow guests to obtain sensitive register content information from other guests.
The isolation provided by the new design with dedicated hypervisors is effective in preventing such guest-to-guest attacks, for there is no shared data or resources between {\sys}s.

\vspace{3mm} \noindent \textbf{Attack from Guest to {\sys}.} 
In the analysis above, we assumed that {\sys} could suffer from all the vulnerabilities of the traditional hypervisors,
but please note that {\sys} is developed in Rust, a high-performance language with guarantees of memory-safe and thread-safe, 
and the ratio of unsafe code is kept below 4.7\%.
This greatly reduces the security risk caused by memory vulnerabilities~\cite{cve201816882, cve20197221, cve202129657, cve20140049, cve20131796} and threading vulnerabilities~\cite{cve202129657, cve20196974, cve20147842}, 
such as the use-after-free vulnerabilities~\cite{cve201816882, cve20197221} in traditional hypervisors developed in C/C++.
In addition, vulnerabilities of the user-level {\sys} can be patched promptly without rebooting the host OS.

\vspace{3mm} \noindent \textbf{Attack from {\sys} to Host Kernel.} 
Although this paper does not consider the original vulnerabilities of the kernel, the newly introduced CP driver cannot be ignored.
A compromised {\sys} process may attack the kernel via the privileged code of the CP driver. 
However, compared to the complex logic and huge code base of traditional hypervisor drivers, the CP driver only has 317 LoC.
Such a tiny code base with simple logic allows the CP driver to be checked more easily for security vulnerabilities.

Furthermore, the DV-Ext interface does not grant {\sys} more capabilities to compromise the host kernel. 
First, although the CP driver allows user-level processes to request physical memory,
it can still effectively isolate them with the help of dynamic physical memory checking mechanisms.
Second, the data-plane register and instructions are configured and restricted by the control-plane registers in the CP driver. 
So they only affect the VM's behavior. Also, the host kernel saves and restores these registers on context switches to avoid affecting other processes.
\section{Performance Evaluation}
\label{sec:eval}

\subsection{Experimental Setup}
\label{subsec:eval_setup}

\begin{figure*}[htbp]
    \begin{minipage}[l]{0.19\textwidth}
		\includegraphics[width=\textwidth]{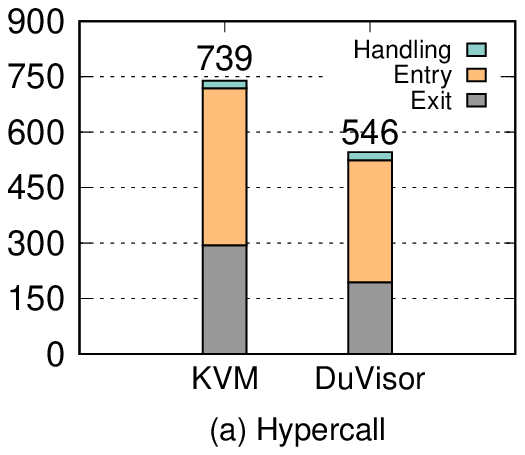}
	\end{minipage}
    \hfill
	\begin{minipage}[l]{0.19\textwidth}
        \includegraphics[width=\textwidth]{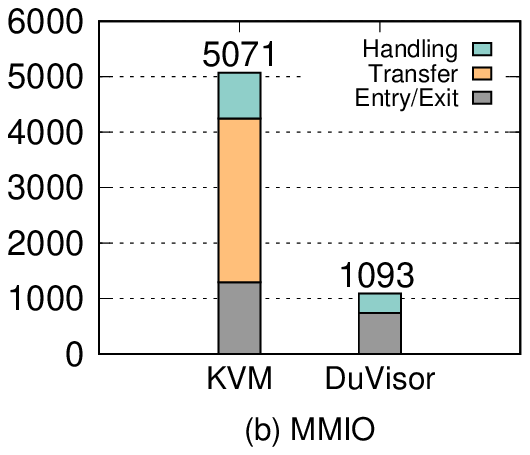}
    \end{minipage}
    \hfill
	\begin{minipage}[l]{0.19\textwidth}
        \includegraphics[width=\textwidth]{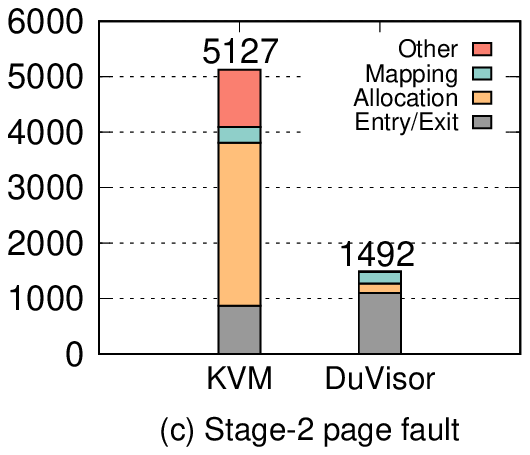}
    \end{minipage}
    \hfill
    \begin{minipage}[l]{0.19\textwidth}
        \includegraphics[width=\textwidth]{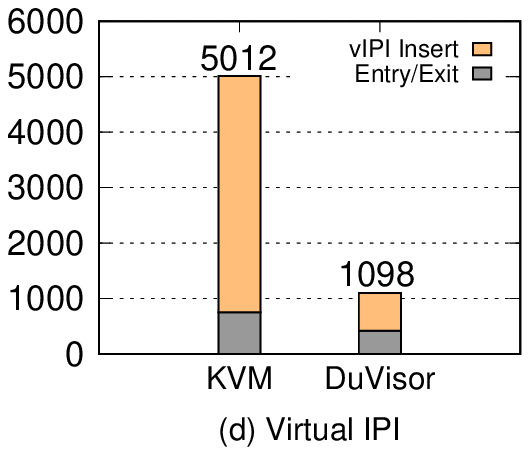}
    \end{minipage}
    \hfill
    \begin{minipage}[l]{0.19\textwidth}
        \includegraphics[width=\textwidth]{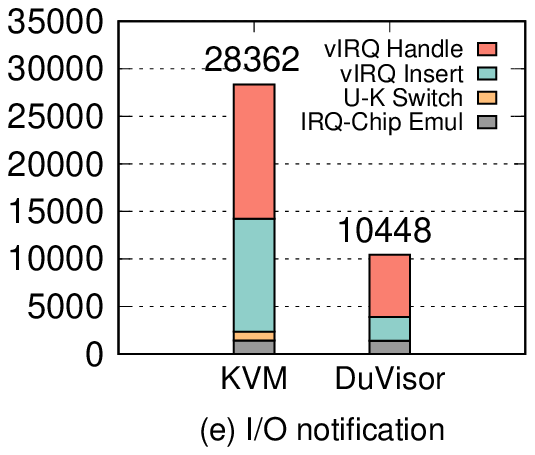}
    \end{minipage}
    \caption{\small{Breakdown of different hypervisor primitives (Unit: cycles).
    \underline{(a) shows a \textit{null} hypercall.}
    \textit{Exit}: from invoking a hypercall in the guest VM to arriving at the hypercall handler in the hypervisor.
    \textit{Entry}: the reverse procedure of \textit{Exit}.
    \textit{Handling}: processing in the hypercall handler.
    \underline{(b) shows an MMIO emulation.}
    \textit{Transfer}: transfers between the kernel in HS-mode and the user-space VMM in HU-mode, which {\sys} gets rid of.
    \textit{Handling}: the emulation of reading data from a virtual console device.
    \textit{Entry/Exit}: from invoking an MMIO operation in the VM to arriving at the MMIO handler and the reverse procedure.
    \underline{(c) shows an S2PF handling.}
    \textit{Allocation}: the memory allocation for the fault GPA.
    \textit{Mapping}: the PTE update in the S2PT.
    \textit{Other}: other logic.
    \underline{(d) shows a virtual IPI sending.}
    \textit{Entry/Exit}: the 1st vCPU's VM exit and the 2nd vCPU's VM entry.
    \textit{vIPI Insert}: the whole process of inserting a virtual interrupt to the 2nd vCPU and
    kicking it by sending IPI (KVM) or UIPI ({\sys}) in the hypervisor.
    \underline{(e) shows an I/O notification during the Netperf UDP test.}
    \textit{IRQ-Chip Emul}: the emulation of the interrupt controller.
    \textit{U-K Switch}: the users-kernel mode switch.
    \textit{vIRQ Insert}: inserting a virtual interrupt and kicking vCPU in the hypervisor.
    \textit{vIRQ Handle}: ACK, processing and EOI a virtual interrupt in the guest VM.
    }}
    \label{fig:micro_benchmark}
\end{figure*}

We run experiments on the cycle-accurate Firesim platform~\cite{karandikar18firesim},
which consists of two FPGA boards.
Each FPGA board has eight RISC-V cores (3.2GHz, rv64imafdc), 16GB RAM and 115GB storage.
For network related benchmarks, we build a local area network (LAN) between
the two boards through 200Gbps IceNICs.
Both FPGA boards are controlled by an EC2 instance that runs CentOS 7.6.1810 on 16-core Intel E5-2686 v4 CPU (2.3GHz) and 240GB RAM.
The firmware for RISC-V is OpenSBI v0.8~\cite{riscv-opensbi}.
The host kernel is Linux kernel 5.10.26 that is equipped with the CP kernel driver to support the {\sys}.
The baseline is KVM~\cite{kvm-riscv} (with H-Ext~\cite{h-ext-pr} support) and \textit{kvmtool} v3.18.0 running on Linux 5.11.0-rc3 (hereinafter called KVM).
Note that kvmtool is the only available hypervisor for KVM on RISC-V platforms,
so that we have to choose it as the performance baseline rather than other hypervisors (e.g., QEMU).

\subsection{Microbenchmarks}
\label{subsec:microbenchmarks}

In this section, we quantify the performance of five frequently used hypervisor primitives.
We leverage the \textit{cycle} CSR to measure CPU cycles.
Fig.~\ref{fig:micro_benchmark} shows the average cost of the five operations in KVM and {\sys}.
We record the total time of repeating each operation 100,000 times and calculate the average cycle count.

In the hypercall microbenchmark, both KVM and {\sys} run a guest VM with a single vCPU pinned to a pCPU.
The guest VM invokes a \textit{null} hypercall, which traps to the hypervisor's handler
and then returns without doing any functional operations.
The hypercall traps to the host kernel in KVM, while trapping to the user space in {\sys}.
The number of cycles between the start of the hypercall and its return position is counted.
As shown in Fig.~\ref{fig:micro_benchmark}(a), {\sys} improves performance by approximately 26.12\% compared with KVM.
Since {\sys} is in user mode, there is no need to perform operations that are only necessary in kernel
(e.g., enabling and disabling preemption and interrupts).
Consequently, {\sys} consumes less time cost during the hypercall handling than KVM.

For S2PF handling, each hypervisor runs a guest VM with a single vCPU.
The guest VM reads one byte from a page unmapped in the S2PT,
resulting in an S2PF.
When compared with KVM, {\sys} achieves about 70.90\% (3,635 cycles) performance improvement.
As is shown in Fig.~\ref{fig:micro_benchmark}(c), we breakdown the S2PF handling process in KVM and find that
the time spent on searching the corresponding \textit{memslot} and memory allocation for the guest fault address
is largest, accounting for about 57.32\% (2,939 cycles) of the total time as the \textit{Allocation} part shows.
The remaining part consists of many generic but cumbersome kernel logic for KVM,
such as finding \textit{virtual memory area (VMA)} and taking locks of \textit{mmap} in Linux,
which accounts for about 20.13\% (1,032 cycles) of the whole process.

For MMIO emulation, a single-vCPU guest VM performs a load operation from an MMIO address region
which traps to the user-level hypervisor and immediately returns the data
from a virtual console device without any additional operations.
We count the elapsed cycles of the MMIO read operation.
The result shows that {\sys} is 78.45\% (3,978 cycles) faster than KVM
due to the shorter path of MMIO handling in {\sys} as in Fig.~\ref{fig:micro_benchmark}(b).
Traditional hypervisors such as KVM offload most MMIO emulations to user mode for security and reliability.
However, such design slows down the MMIO handling for the longer path,
which significantly reduces the efficiency of guest's operations on the virtual devices, such as the network and block devices.
The breakdown shows that 58.17\% (2,950 cycles) of the time cost during the MMIO emulation in KVM
is spent on traveling across the kernel mode.

For virtual IPI, both KVM and {\sys} run a dual-vCPU guest VM and pin two vCPUs to separate cores.
The first vCPU sends a virtual IPI to the second vCPU and then waits for the completion response.
When the second vCPU receives the virtual IPI, it notifies the first vCPU by sending a virtual IPI back.
We collect the cycles between issuing the virtual IPI and receiving the completion on the first vCPU.
{\sys} is 78.09\% (3,914 cycles) faster than KVM.
According to the breakdown, the acceleration mainly comes from the faster UIPI as in Fig.~\ref{fig:micro_benchmark}(d).
A virtual IPI issued by the guest VM is trapped into the host kernel for emulation.
If the receiver vCPU is executing on a separate physical CPU core,
the hypervisor will kick it by sending a hardware IPI to that core.
To send an IPI, {\sys} executes the HUSUIPI instruction in HU-mode without any mode switches.
Nonetheless, the host kernel (HS-mode) relies on the firmware (M-mode) to issue an IPI,
which in the RISC-V KVM implementation includes an ECALL (environment call, the instruction which causes a software trap to a more privileged mode on RISC-V~\cite{riscv-privileged}) to the firmware.
By removing mode switches, {\sys} achieves a considerable reduction in virtual IPI latency compared with KVM.

For I/O notification, we breakdown the RX process of network packets from the backend driver to
the interrupt handler of the single-vCPU guest VM during the Netperf UDP latency test.
The vCPU and the RX I/O thread are pinned to different pCPUs.
As shown in Fig.~\ref{fig:micro_benchmark}(e), {\sys} is 63.16\% (17,914 cycles) faster than KVM:
The UIPI-based user-level notification in {\sys} boosts the virtual interrupt inserting
by 78.92\% (9,371 cycles) compared with KVM.
Meanwhile, processing a virtual interrupt in the guest VM in {\sys} costs 53.6\% (7,579 cycles) less than KVM
because of the fact that accessing the virtual interrupt controller on current RISC-V platforms
incurs MMIO VM exits to the user-space hypervisor.

\subsection{Real-world Applications Performance}
\label{subsec:eval_apps}

\begin{table}[htp]
    \centering
    \scriptsize{
    \begin{tabularx}{0.45\textwidth}{|l|X|}
    \hline
    \textbf{Name} & \textbf{Description} \\
    \hline 
    Netperf   & Netserver v2.6.0 on the local server (guest VM) and Netperf v2.6.0 on the remote client (native) to test the UDP latency for 5 seconds.  \\
    \hline
    iperf3    & iperf v3.9 on both the local server (guest VM) and the remote client (native) to test the TCP throughput for 10 seconds. \\
    \hline 
    Memcached & Memcached v1.6.10 running the memtier benchmark 1.3.0 on the remote client to test transactions per second. The thread number is set to the same as the number of server vCPUs. Each round of test lasts 5 seconds. \\
    \hline
    FileIO    & FileIO test in sysbench v0.4.12 with 4 threads concurrently and 512MB file size in random read/write mode. \\
    \hline
    Hackbench & Hackbench using Unix domain sockets and default 10 process groups running in 100 loops, measuring the time cost. \\
    \hline 
    Untar     & Untar extracting the lmbench v2 tarball using the standard tar utility, measuring the time cost. \\
    \hline
    CPU-Prime & CPU test in sysbench v0.4.12 that calculates prime numbers up to the max prime 10000. The thread number is set to the same as the number of server vCPUs. \\
    \hline
    \end{tabularx}
    }
    \caption{\small{Descriptions of application benchmarks.}}
    \label{tab:app-benchmark}
\end{table}

In this section, we evaluate the performance of a variety of real-world applications,
compare the results between KVM and {\sys}, and analyze the reasons for these performance differences.
Table~\ref{tab:app-benchmark} lists applications we used for benchmark.
We measure the performance in four types of guest VMs with 1, 2, 4 and 6 vCPUs.
All VMs are equipped with 512MB memory, virtio-based network and block devices.
For single-VM tests, we enable the kernel-bypass NIC driver in the network backend for both KVM and {\sys}.
Each vCPU of a guest VM is pinned to a separate physical CPU core to
avoid potential instability caused by the host kernel scheduler.
The kernel-bypass NIC driver dedicates two CPUs for RX and TX I/O threads
respectively to maximize network performance.
In multi-VM tests, we still use the kernel network stack because
our current implementation of the kernel-bypass NIC driver does not support network switching.
If the total number of vCPU and I/O threads exceeds eight,
we first ensure that vCPUs are on different physical cores,
while I/O threads may co-locate with vCPUs.

\begin{figure*}[htbp]
    \begin{minipage}[l]{0.24\textwidth}
		\includegraphics[width=\textwidth]{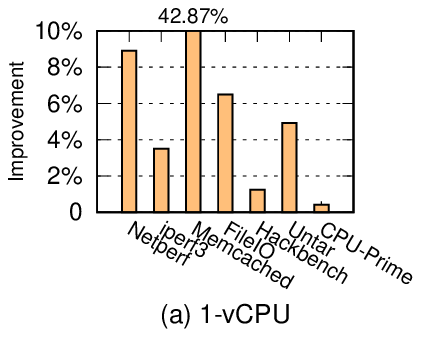}
	\end{minipage}
    \hfill
	\begin{minipage}[l]{0.24\textwidth}
        \includegraphics[width=\textwidth]{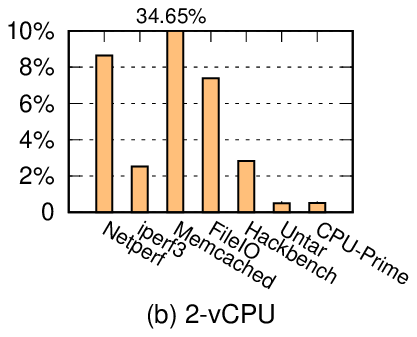}
    \end{minipage}
    \hfill
	\begin{minipage}[l]{0.24\textwidth}
        \includegraphics[width=\textwidth]{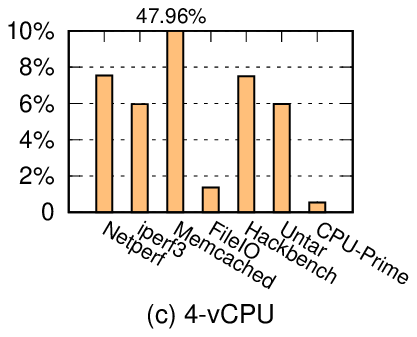}
    \end{minipage}
    \hfill
	\begin{minipage}[l]{0.24\textwidth}
        \includegraphics[width=\textwidth]{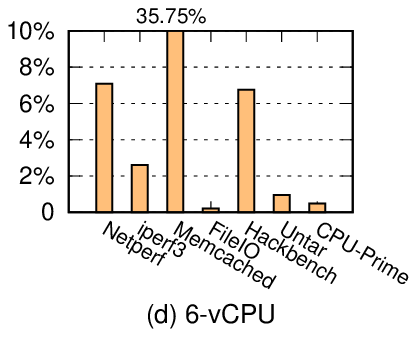}
    \end{minipage}
    \caption{\small{Normalized performance of real-world applications of {\sys} compared with KVM.
    The Y-axis indicates the performance improvement of {\sys} over KVM.}}
    \label{fig:app-benchmark}
\end{figure*}

\begin{figure*}[htbp]
    \begin{minipage}[l]{0.24\textwidth}
		\includegraphics[width=\textwidth]{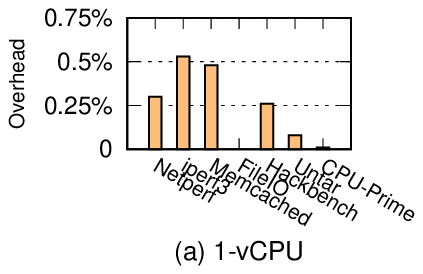}
	\end{minipage}
    \hfill
	\begin{minipage}[l]{0.24\textwidth}
        \includegraphics[width=\textwidth]{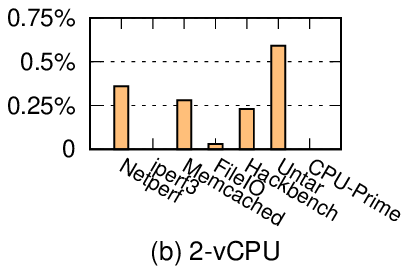}
    \end{minipage}
    \hfill
	\begin{minipage}[l]{0.24\textwidth}
        \includegraphics[width=\textwidth]{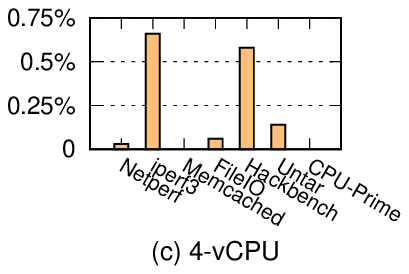}
    \end{minipage}
    \hfill
	\begin{minipage}[l]{0.24\textwidth}
        \includegraphics[width=\textwidth]{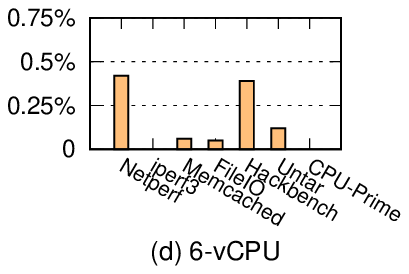}
    \end{minipage}
    \caption{\small{Normalized performance of real-world applications of KVM with context switch logic of DV-Ext compared with vanilla KVM.
    The Y-axis presents the overhead brought by the context switch logic of DV-Ext.}}
    \label{fig:impact-on-kvm}
\end{figure*}

As shown in Fig.~\ref{fig:app-benchmark}(a),
all applications in an 1-vCPU guest VM in {\sys} have better performance than KVM.
Memcached, a network-intensive application, shows about 43\% improvement.
According to our microbenchmark, the major reason for this improvement is that 
the network I/O notification in {\sys} costs merely one-third the price of KVM.
We further gather the time cost of VM exits during the Memcached benchmark:
VM exits caused by MMIO are of the highest percentage in both KVM (58.64\%) and {\sys} (66.34\%).
The time cost of {\sys}'s MMIO handling time cost is merely 35.87\% of KVM,
resulting in the large performance improvement.
But we believe that the {\sys} architecture will not deliver such huge performance gains
on platforms that provide hardware interrupt virtualization (e.g., ARM vGIC) to
reduce excessive MMIO VM exits.
Other I/O-intensive applications have less than 10\% improvements:
FileIO and iperf3 benchmarks aim to saturate the I/O bandwidth, and the virtio optimizes
such scenario by reducing the frequency of I/O notifications.
Therefore, the less I/O notification is the reason why their performance improvements are smaller than Memcached.
Netperf evaluates the network latency in a ping-pong method between the client and the server.
Though {\sys}'s I/O notification is much faster,
the time cost of network stacks in both client and server dominates the whole procedure
(network stacks cost 50 \textmu s out of the total 60 \textmu s), lowering the performance improvement.

Fig.~\ref{fig:app-benchmark}(b) shows the performance of a 2-vCPU guest VM,
{\sys} achieves up to 35\% improvement over KVM in the Memcached benchmark.
In addition to {\sys}'s faster I/O notifications,
the more efficient virtual IPI delivery in {\sys} is another factor of the improvement.
In a multithreaded application, one thread needs to perform operations
such as event notifications and message passing to communicate with other threads.
When two threads that require communications are running on different vCPUs,
one must send a virtual IPI in the guest VM to notify the other side.

Fig.~\ref{fig:app-benchmark}(c) and (d) shows the performance of 4-vCPU and 6-vCPU guest VMs,
similar to the 2-vCPU case, {\sys} outperforms KVM by up to 48\% and 36\%.

\textbf{Performance Impact on KVM VMs:} To investigate the performance impact of {\sys}'s host kernel modifications and DV-Ext on KVM VMs,
we evaluate the performance of all real-world applications in KVM VMs after
porting the modifications to the host kernel and running it on the DV-Ext enabled hardware.
The results in Fig.~\ref{fig:impact-on-kvm}(a), (b), (c) and (d) show that 
there is no discernible performance degradation when compared with unmodified KVM VMs,
demonstrating that the impact of supporting DV-Ext in the host kernel is negligible.

\subsection{Scalability}
\label{subsec:eval_scalability}

\begin{figure}[htbp]
    \begin{minipage}[l]{0.15\textwidth}
		\includegraphics[width=\textwidth]{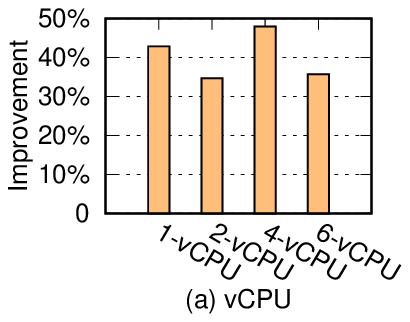}
	\end{minipage}
    \hfill
	\begin{minipage}[l]{0.15\textwidth}
        \includegraphics[width=\textwidth]{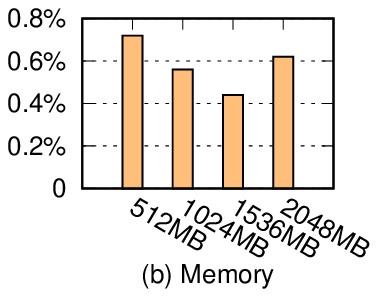}
    \end{minipage}
    \hfill
	\begin{minipage}[l]{0.15\textwidth}
        \includegraphics[width=\textwidth]{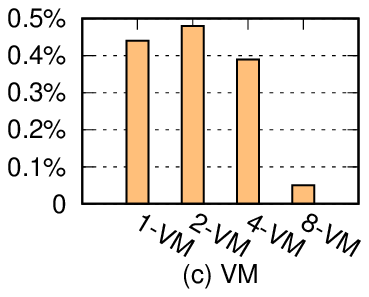}
    \end{minipage}
    \caption{\small{The evaluation of {\sys}'s scalability using Memcached compared with KVM.
    (a) and (b) show {\sys}'s normalized improvement of a different numbers of vCPUs and sizes of memory respectively.}}
    \label{fig:scalability}
\end{figure}

\noindent\textbf{Scaling vCPU number:} 
To show {\sys}'s vCPU scalability compared with KVM, we choose Memcached as the benchmark
because its multi-threading model is able to benefit from multiple vCPUs.
We test Memcached in a guest VM with 512MB memory and
increase its vCPU number from 1 to 2, 4 and 6.
The result is shown in Fig.~\ref{fig:scalability}(a).
{\sys} keeps outperforming KVM in all cases from 34.65\% to 47.96\% and
thus scales well as the number of vCPUs grows.

\noindent\textbf{Scaling Memory:} 
To show {\sys}'s memory scalability compared with KVM,
we run STREAM~\cite{mccalpin2006stream}, a memory-intensive benchmark,
in a 4-vCPU guest VM with 512MB, 1024MB, 1536MB and 2048MB memory.
As shown in Fig.~\ref{fig:scalability}(b), 
{\sys} achieves slightly better performance than KVM in all cases.
Compared with the memory virtualization in KVM, {\sys} enables the HU-mode to configure the S2PT
via extending the PMP checking with minor modifications,
so that there is little impact on the memory access latency and thus scales well as the memory size grows.

\noindent\textbf{Scaling VM number:}
To reflect the scalability of {\sys} in the situation of multiple VMs,
we evaluate the performance of different numbers of guest VMs in {\sys} and KVM.
We test CPU-Prime as a CPU-intensive workload in 1, 2, 4 and 8 VMs concurrently.
As shown in Fig.~\ref{fig:scalability}(c), {\sys} has good scalability and performs as well as KVM.

\subsection{PMP Checking}
\label{subsec:pmp_checking}

{\sys} slightly extends the PMP hardware with a V bit to verify the validity of memory accesses from guest VMs.
We set up a guest VM with 4-vCPU and 1024MB memory to 
evaluate and compare performance with and without PMP V bit hardware to
see if this checking mechanism imposes additional memory access overhead on VMs.
We utilize Memcached as the memory-intensive benchmark and allocate 900MB of memory to it.
The results reveal that the performance of the VM with and without PMP checking has little difference. 
As a result, the physical memory restriction imposed by {\sys} has invisible effect on VM performance.
\section{Discussion}
\label{sec:discuss}



\vspace{3mm} \noindent \textbf{Virtual Machine Management.}
All VM states are completely managed by {\sys}, so that it can support normal VM management operations.
First, cloud operators can suspend the VM at any time to make a VM checkpoint and restore it from this image afterwards.
Second, different intrusion detection systems (IDS)~\cite{roschke2009intrusion, dunlap2002revirt} and virtual machine introspection (VMI)~\cite{garfinkel2003virtual} tools can be implemented in {\sys} to monitor VM behaviors.
Moreover, {\sys} can support different live VM migration strategies~\cite{clark2005live, adam2018vm} since it is able to suspend a VM and keep copying its dirty pages to a remote instance.

\vspace{3mm} \noindent \textbf{Rapid Development and Deployment.}
Delegated virtualization enables developers to leverage rich development resources and debugging tools in user space 
to quickly react to various security vulnerabilities, rapidly evolving hardware features and application requirements.
Besides, a user-level hypervisor accelerates the hypervisor upgrade cycle since developers can replace the old hypervisor without rebooting the host kernel.

\vspace{3mm} \noindent \textbf{Nested Virtualization.} 
The existing nested virtualization~\cite{ben2010turtles,zhang2011cloudvisor} mandates that all VMs exits have to be intercepted by L0 hypervisor (bare-metal one) before being handled by 
L1 hypervisor (nested one), which incurs tremendous runtime overhead. More levels of nested virtualization further increase the overhead
exponentially.
In the future, we plan to extend the idea of delegated VM exits (DVE) to optimize nested virtualization, which 
allows the VM exits to trap directly to the L1 hypervisor that has the information to handle them without 
involving the L0 one.





\section{Related Work}
\label{sec:related}
This section compares {\sys} with closely related work, including approaches that protects VMs, reduces virtualization 
overhead, deprivileges kernel features to userspace, and design new hardware for systems.

\subsection{Securing VMs}
Many studies have considered how to achieve better isolation for VMs with unreliable hypervisors to mitigate security threats. 
One solution is
to propose hardware extensions to remove the vulnerable hypervisor out of
TCB~\cite{szefer2012architectural, azab2010hypersentry, jin2011architectural, keller2010nohype}.
In particular, NoHype~\cite{keller2010nohype} proposes to completely remove the hypervisor to reduce the attack surface and avoid 
VM exits. It dedicates physical cores, memory, I/O devices by implementing these 
virtualization functions in hardware chips, which disallows resource oversubscription and is thus inviable for practical deployment. 
Different from these work, {\sys} puts virtualization functions to user mode to minimize the runtime attack surface of a hypervisor. 
Even if a malicious VM takes over its own hypervisor, it cannot make direct attacks on other VMs, including DoS attacks.


Existing studies also have explored how to defend VMs via software methods~\cite{li2019protecting, colp2011breaking, steinberg2010nova, shi2017deconstructing, wang2012isolating, wang2010hypersafe, wang2012isolating, dahlin2011toward}.
NOVA~\cite{steinberg2010nova} builds a microhypervisor based on the microkernel architecture.
For KVM that is widely deployed in commercial scenarios, DeHype~\cite{wu2013taming} tries to move most parts of KVM into user mode. However, it still needs a HypLet in kernel mode since 
the sensitive instructions of virtualization can only be used in this mode. Such design incurs 
a large number of system calls and thus runs slower than KVM, which has already brought non-trivial performance overhead
due to excessive ring crossings.
For Xen~\cite{barham2003xen} hypervisor, Nexen~\cite{shi2017deconstructing} deconstructs it into
different instances of non-privileged service slices. 
HypSec~\cite{li2019protecting} decouples the hypervisor into a small corevisor and untrusted hypervisor via exploiting ARM TrustZone and virtualization extension.
All these software approaches leverage the traditional interface of hardware virtualization, 
{\sys} proposes a novel hardware interface and downgrades the whole hypervisor to user space while also reducing world switches.


\subsection{Reducing Virtualization Overhead}
One major cause of virtualization overhead is costly world switches~\cite{zhang2020high, humphries2021case, mi2020mostly}.
To this end, existing approaches try to reduce the number of world switches~\cite{agesen2012software, dong2011optimizing, ahmad2011vic}.
To boost the interrupt virtualization, ELI~\cite{gordon2012eli} passes interrupts directly to guest VMs without the involvement of the hypervisor.
CloudVisor-D~\cite{mi2020mostly} leverages an Intel instruction (VMFUNC) to allow a VM to directly interact with the hypervisor 
without trapping into the privileged mode, which effectively reduces the number of world switches. 
But it can only support nested virtualization and needs to modify the guest OS to proactively invoke the VMFUNC instruction.
BM-Hive~\cite{zhang2020high} offers a physically isolated machine to run a VM via bare-metal virtualization, totally avoiding
the virtualization overhead. 
In contrast, {\sys} reduces the virtualization overhead by hardware-software co-design and removing the kernel host from the path of VM exit handling.

Memory translation is another source of virtualization overhead since it may lead to at most 24 memory accesses for walking
two page tables~\cite{gandhi2016agile}. Solutions have been proposed either to increase
the number of TLB entries~\cite{park2017hybrid, ryoo2017rethinking} or use huge pages to increase the TLB hit rate~\cite{pham2015large}.
All these optimization techniques can be applied to {\sys} to further improve the virtualization performance.

\subsection{Moving Kernel Functions to Userspace}
Deprivileging kernel features to user space is a classic approach to ease kernel development or improve system reliability.
Microkernels are one typical design~\cite{liedtke1995on, klein2009sel4, ganapathy2008design}, where system services such as file systems and drivers run in user mode. 
Therefore, costly inter-process communication (IPC) is frequently used to connect these services~\cite{klein2009sel4}.
For monolithic kernels, similar methods also exist, which implement the file systems~\cite{miller2021high, dong2019performance}, scheduler~\cite{humphries2021ghost}, network service~\cite{marty2019snap} in user space.
Different from these related work, {\sys} is the first system that fully moves runtime virtualization functions to user space, which benefits hypervisors running in both monolithic kernels~\cite{kivity2007kvm, barham2003xen} and microkernels~\cite{steinberg2010nova, heiser2010okl4}.

\subsection{Hardware-Software Co-design}
There have also been many research efforts on improving systems by designing new hardware~\cite{schrammel2020donky, vilanova2017direct, dong2019xpc, lim2017neve, dall2016arm, lim2017neve}.
Donky~\cite{schrammel2020donky} implements an enhanced version of the Intel MPK mechanism via modifications to the N-Ext of RISC-V.
XPC~\cite{dong2019xpc} and dIPC~\cite{vilanova2017direct} boost IPC performance via the new architectures and interfaces of hardware.
Most importantly, Dall et al.~\cite{dall2016arm} and NEVE~\cite{lim2017neve} propose a set of architectural changes to existing ARM hardware virtualization.
Though {\sys} also overcomes the design flaws of traditional hypervisors via hardware-software co-design, it does not target a specific hardware platform.
\section{Conclusion}
\label{sec:concl}


This paper presents delegated virtualization via retrofitting existing hardware with a new delegated virtualization extension. 
We build a user-level hypervisor atop the hardware extension, called {\sys}, that directly handles all VM operations without 
trapping into the kernel at runtime.  
Experiments results show that {\sys} outperforms traditional hypervisors. 

{
\small{
\bibliographystyle{plain}
\bibliography{ms}
}

\end{document}